\documentclass[3p,times,preprint,twocolumn]{elsarticle}
\biboptions{sort&compress}
\usepackage{enumerate}
\usepackage{mathtools}
\usepackage{color}

\usepackage[normalem]{ulem}

\makeatletter

\usepackage{amssymb,amsmath,amsthm}
\usepackage[title,toc]{appendix}
\usepackage{graphicx}
\usepackage{color}
\usepackage{hyperref}
\theoremstyle{definition}

\begin{document}

\begin{frontmatter}

\title{Separability of the Klein-Gordon equation for rotating spacetimes obtained from Newman-Janis algorithm}

\author[1,2]{Che-Yu Chen}
\ead{b97202056@gmail.com}

\author[1,2,3]{Pisin Chen}
\ead{pisinchen@phys.ntu.edu.tw}

\address[1]{Department of Physics and Center for Theoretical Sciences, National Taiwan University, Taipei, Taiwan 10617}
\address[2]{LeCosPA, National Taiwan University, Taipei, Taiwan 10617}
\address[3] {Kavli Institute for Particle Astrophysics and Cosmology, SLAC National Accelerator Laboratory, Stanford University, Stanford, CA 94305, U.S.A.}

\begin{abstract}
In the literature, the Newman-Janis algorithm (NJA) has been widely used to construct stationary and axisymmetric spacetimes to describe rotating black holes. In addition, it has been recently shown that the general stationary and axisymmetric spacetime generated through NJA allows the complete separability of the null geodesic equations. In fact, the Hamilton-Jacobi equation in this spacetime is also separable if one of the metric functions is additively separable. In this work, we further study the conditions for a separable Klein-Gordon equation in such a general spacetime. The relations between the NJA spacetime and other parameterized axially symmetric spacetimes in the literature are also discussed. 
\end{abstract}

\end{frontmatter}

\section{Introduction}

In Einstein's general relativity (GR), it is well-known that isolated and rotating black holes are described by the Kerr metric, which is axisymmetric and asymptotically flat. The Kerr metric is a solution to the vacuum Einstein equation and, according to the no-hair theorem, it is completely described by two external observable classical parameters: its mass and angular momentum. However, so far there is still no solid proof that the no-hair theorem must be true. Actually, any amount of violations of the no-hair theorem, no matter how tiny it would be, may imply the necessity to modify GR.   

In order to examine whether the no-hair theorem is indeed valid, several strong-field tests have been suggested, such as using gravitational waves associated with black hole perturbations, extreme mass-ratio inspirals (EMRIs), and black hole shadows \cite{Berti:2015itd}. These strong-field tests can be implemented in a theory-agnostic way, but this relies on the construction of a modified spacetime deviating from the Kerr metric where those deviations can be expressed parametrically. In the literature, there have been several parameterized Kerr-like spacetimes being proposed and studied. See Refs.~\cite{Glampedakis:2005cf,Johannsen:2011dh,Johannsen:2015pca,Cardoso:2014rha,Cardoso:2015xtj,Konoplya:2016jvv,Ghasemi-Nodehi:2016wao,Konoplya:2018arm,Johannsen:2015qca,Ghasemi-Nodehi:2015raa,Younsi:2016azx,Ni:2016uik,Glampedakis:2017dvb} and references therein.

In addition to the aforementioned parametric models, the Newman-Janis algorithm (NJA) is another powerful tool to construct stationary and axisymmetric spacetimes \cite{Newman:1965tw,Newman:1965my,Drake:1998gf}. Starting with a static and spherically symmetric seed metric, one can use NJA to generate the corresponding stationary and axisymmetric spacetime. This approach not only works well for the Kerr and Kerr-Newman metrics, it is also widely used to construct effective models for rotating black holes in different theories \cite{Modesto:2010rv,Azreg-Ainou:2014pra,Bambi:2013ufa,Shaikh:2019fpu,Tsukamoto:2017fxq,Kumar:2017qws,Xu:2018wow,Rahman:2018fgy}. In fact, since the derivation of rotating metrics in generally modified theories of gravity could be much more difficult than that of spherically symmetric metrics, NJA is also commonly used to investigate rotating black holes in modified theories of gravity.   

Even though NJA is powerful and easy to implement, it still has some shortcomings. For example, it has been proven that the rotating black holes generated through NJA in some modified theories of gravity are actually not the solution of the same theory \cite{Hansen:2013owa,Kumar:2019uwi}. Furthermore, there are some ambiguities in the whole process of NJA \cite{Erbin:2016lzq}, such as those in the complexification of metric functions, and the uniqueness of the complex coordinate transformations. However, NJA can also be regarded as a method to construct parametrized Kerr-like spacetimes because the resultant metric may depend on some arbitrary functions which should be fixed via observational tests. 

In order to treat the NJA spacetime as a parametrized Kerr-like spacetime, one has to first investigate the properties of such a resultant metric. In Ref.~\cite{Shaikh:2019fpu}, it has been proven that the general stationary and axisymmetric metric generated via NJA allows for completely separable null geodesic equations. In this work, we will study the separability of the Klein-Gordon equation for such a general spacetime derived via NJA (we shall call it NJA metric from now on). The separability of the Klein-Gordon equation \cite{Brill:1972xj} is crucial in dealing with several physical problems, such as the quasinormal modes and the scattering problem of the black hole. We will show explicitly what conditions would this class of spacetimes actually count on to render the separability of the Klein-Gordon equation. Then, we will compare the NJA metric with other parameterized Kerr-like spacetimes in the literature, which were proposed in Refs.~\cite{Johannsen:2015pca}, \cite{Konoplya:2018arm}, \cite{Papadopoulos:2018nvd}, and \cite{Carter:1968ks}, respectively. They are characterized by the existence of the Carter constant of motion and the separability of the Hamilton-Jacobi equation. The metric proposed in \cite{Konoplya:2018arm}, especially, allows for the separability of the Hamilton-Jacobi and the Klein-Gordon equations. We will illustrate the relations between these parameterized metrics.

The rest of this paper is organized as follows. In Section~\ref{sec.2}, we review the general setup of NJA and the generation of the general stationary and axisymmetric spacetime. Section~\ref{sec.3} presents the condition for the separability of the Klein-Gordon equation in such a general spacetime. In Section~\ref{sec.4}, we compare the NJA metric with other parameterized Kerr-like spacetimes in the literature. We draw our conclusions and discussions in Section~\ref{sec.5}.

\section{The general Newman-Janis Algorithm}\label{sec.2}
In this section, we review how NJA works to generate stationary and axisymmetric spacetimes. We start with the most general static and spherically symmetric spacetime as a seed metric. This seed metric can be written as
\begin{equation}
ds^2=-g(r)dt^2+\frac{dr^2}{f(r)}+h(r)d\Omega^2\,,\label{SSSmetric}
\end{equation}
where $f(r)$, $g(r)$ and $h(r)$ are functions of the radial coordinate $r$. The first step of NJA is to introduce the advanced null coordinates ($u,r,\theta,\phi$), in which $u$ is defined by
\begin{equation}
u\equiv t-r_*\,,
\end{equation}
where $r_*$ is the tortoise radius satisfying
\begin{equation}
\frac{dr_*}{dr}\equiv \frac{1}{\sqrt{fg}}\,.
\end{equation}
In the advanced null coordinates, the metric \eqref{SSSmetric} reads
\begin{equation}
ds^2=-g(r)du^2-2\sqrt{\frac{g(r)}{f(r)}}dudr+h(r)d\Omega^2\,.\label{SSSanc}
\end{equation}

Then, we express the inverse metric $g^{\mu\nu}$ of the spacetime \eqref{SSSanc} by using a null tetrad $Z^\mu_a=(l^\mu,n^\mu,m^\mu,\bar{m}^\mu)$ such that
\begin{equation}
g^{\mu\nu}=-l^\mu n^\nu-l^\nu n^\mu+m^\mu\bar{m}^\nu+m^\nu\bar{m}^\mu\,,\label{metrictetradnull}
\end{equation}
where $\bar{m}^\mu$ is the complex conjugate of $m^\mu$. Given the metric \eqref{SSSanc}, the components of the null tetrad $Z^\mu_a$ can be written as
\begin{align}
l^\mu&=\delta_r^\mu\,,\qquad n^\mu=\sqrt{\frac{f(r)}{g(r)}}\delta_u^\mu-\frac{f(r)}{2}\delta_r^\mu\,,\\
m^\mu&=\frac{1}{\sqrt{2h(r)}}\left(\delta_\theta^\mu+\frac{i}{\sin\theta}\delta_\phi^\mu\right)\,.
\end{align}
Note that these tetrad components satisfy the following relations
\begin{align}
l_\mu l^\mu&=n_\mu n^\mu=m_\mu m^\mu=l_\mu m^\mu =n_\mu m^\mu=0\,,\\
l_\mu n^\mu&=-1\,,\qquad m_\mu \bar{m}^\mu=1\,.
\end{align}

Next, we introduce the following complex coordinate transformation
\begin{equation}
u'=u-ia\cos\theta\,,\qquad r'=r+ia\cos\theta\,,
\end{equation}
where $a$ is a constant and it can be treated as the spin parameter of the final rotating black hole. Note that the $\theta$ and $\phi$ coordinates remain unchanged, and the new coordinates $u'$ and $r'$ are both real-valued. After the coordinate transformation, the metric functions are not functions of $r'$ only. In fact, they should be functions of the coordinate $\theta$ as well, depending on how one complexifies the $r'$ coordinate. In this work, we will skip the complexification procedure as was first introduced in Ref.~\cite{Azreg-Ainou:2014pra}. In this regard, we will treat the metric functions after complexifications as some arbitrary functions of $r'$ and $\theta$ for the time being: 
\begin{equation}
f(r)\rightarrow F(r',\theta)\,,\quad g(r)\rightarrow G(r',\theta)\,,\quad h(r)\rightarrow H(r',\theta)\,.
\end{equation}
See also Ref.~\cite{Modesto:2010rv} for another way of complexification. 

According to the coordinate transformation rule ${Z'}_a^\mu=(\partial x'^\mu/\partial x^\nu)Z_a^\nu$, the null tetrad expressed in the new coordinate system can be written as
\begin{align}
l'^\mu&=\delta_{r'}^\mu\,,\qquad n'^\mu=\sqrt{\frac{F(r',\theta)}{G(r',\theta)}}\delta_{u'}^\mu-\frac{F(r',\theta)}{2}\delta_{r'}^\mu\,,\\
m'^\mu&=\frac{1}{\sqrt{2H(r',\theta)}}\left[ia\sin\theta\left(\delta_{u'}^\mu-\delta_{r'}^\mu\right)+\delta_\theta^\mu+\frac{i}{\sin\theta}\delta_\phi^\mu\right]\,.
\end{align}
From now on, we will drop the prime for the sake of simplicity. Considering the new tetrad and using Eq.~\eqref{metrictetradnull}, one can build the following line element:
\begin{align}
ds^2=&-Gdu^2-2\sqrt{\frac{G}{F}}dudr\nonumber\\&+2a\sin^2\theta\left(G-\sqrt{\frac{G}{F}}\right)dud\phi\nonumber\\
&+2a\sqrt{\frac{G}{F}}\sin^2\theta drd\phi+Hd\theta^2\nonumber\\
&+\sin^2\theta\left[H+a^2\sin^2\theta\left(2\sqrt{\frac{G}{F}}-G\right)\right]d\phi^2\,.\label{metricrotatingur}
\end{align}

The last step of NJA is to rewrite the metric \eqref{metricrotatingur} in the Boyer-Lindquist coordinate ($t,r,\theta,\psi$) such that its $g_{t\psi}$ component is the only off-diagonal component. In general, this can be achieved by introducing the following transformations:
\begin{equation}
du=dt+a_1(r)dr\,,\qquad d\phi=d\psi+a_2(r)dr\,,\label{a1a2trans}
\end{equation}
where
\begin{align}
a_1(r)=-\frac{X(r)}{\Delta(r)}\,,\\
a_2(r)=-\frac{a}{\Delta(r)}\,,
\end{align}
and
\begin{align}
\Delta(r)\equiv F(r,\theta)H(r,\theta)+a^2\sin^2\theta\,,\nonumber\\ X(r)\equiv \sqrt{\frac{F(r,\theta)}{G(r,\theta)}}H(r,\theta)+a^2\sin^2\theta\,.\label{DeltaX}
\end{align}
It is important to emphasize that $a_1(r)$ and $a_2(r)$ are functions of $r$ only. They cannot depend on $\theta$ in order to retain the integrability of the transformations \eqref{a1a2trans} \cite{AzregAinou:2011fq,Bambi:2013ufa}. This means that $\Delta(r)$ and $X(r)$ should also depend on $r$ only, even though the metric functions may contain arbitrary dependence on $\theta$. Note that the condition for integrability of such transformations was first pointed out in Ref.~\cite{AzregAinou:2011fq}.

It should also be stressed that even though the NJA approach we consider here and that in Ref.~\cite{Azreg-Ainou:2014pra} are very similar, there is actually a subtle difference. At the end of the algorithm in Ref.~\cite{Azreg-Ainou:2014pra}, the author linked the new metric functions ($F(r,\theta)$, $G(r,\theta)$, and $H(r,\theta)$) to the seed metric in a certain way (Eq.~(14) in Ref.~\cite{Azreg-Ainou:2014pra}), such that the integrability condition of Eq.~\eqref{a1a2trans} is guaranteed. However, in this paper we have not assumed any relation between the new metric functions and the seed metric functions. The new metric functions are assumed to be as general as possible, as long as the integrability condition is satisfied.

Finally, the rotating metric in the Boyer-Lindquist coordinate can be written as
\begin{align}
ds^2=&-Gdt^2+2a\sin^2\theta\left(G-\sqrt{\frac{G}{F}}\right)dtd\psi\nonumber\\&+Hd\theta^2+\frac{H}{\Delta}dr^2\nonumber\\
&+\sin^2\theta\left[H+a^2\sin^2\theta\left(2\sqrt{\frac{G}{F}}-G\right)\right]d\psi^2\,,\label{NJAmetric}
\end{align}
subject to the constraint that $\Delta(r)$ and $X(r)$ given in Eq.~\eqref{DeltaX} should depend on $r$ only. It should be emphasized that we have not made any specific assumption on the seed metric at the beginning. In addition, we have not considered any particular way of complexification in the algorithm. The only requirement is the integrability of the transformations \eqref{a1a2trans}, which implies that $\Delta(r)$ and $X(r)$ only depend on $r$. Therefore, the metric \eqref{NJAmetric} is the most general rotating spacetime metric based on the applicability of NJA. In Ref.~\cite{Shaikh:2019fpu}, it has been shown that the general spacetime \eqref{NJAmetric} allows for the separability of the null geodesic equations. However, this does not imply that the Hamilton-Jacobi equation is also separable. In fact, the separability of the Hamilton-Jacobi equation of this metric requires that the metric function $H(r,\theta)$ should be additively separable. For the allowance of separable null geodesic equations, this requirement is not necessary.

Note that the Kerr metric is recovered when 
\begin{equation}
F=G=1-\frac{2Mr}{H}\,,\qquad H=r^2+a^2\cos^2\theta\,.
\end{equation}
In this case, from Eq.~\eqref{DeltaX}, we obtain
\begin{equation}
\Delta(r)=r^2-2Mr+a^2\,,\qquad X(r)=r^2+a^2\,.
\end{equation}

\section{Separation of variables in the Klein-Gordon equation}\label{sec.3}
The Klein-Gordon equation describes the evolution of a massive scalar field $\Phi$ on curved spacetimes. It is given by
\begin{align}
&\Box\Phi-\mu^2\Phi\nonumber\\=&\,\frac{1}{\sqrt{-g}}\partial_\mu\left(g^{\mu\nu}\sqrt{-g}\partial_\nu\Phi\right)-\mu^2\Phi=0\,,\label{KGeq}
\end{align} 
where $\mu$ stands for the mass of the scalar field. Inserting the metric \eqref{NJAmetric} into Eq.~\eqref{KGeq} and using Eq.~\eqref{DeltaX}, the Klein-Gordon equation can be written as
\begin{align}
0&=\partial_r\left(\Delta Y\partial_r\Phi\right)-\frac{a^2}{\Delta}Y\partial_\psi^2\Phi+\frac{2aY}{\Delta}\left(\Delta-X\right)\partial_t\partial_\psi\Phi\nonumber\\
&-Y^2\left(X-a^2\sin^2\theta\right)\mu^2\Phi-\frac{Y}{\Delta}X^2\partial_t^2\Phi\nonumber\\
&+\frac{1}{\sin\theta}\partial_\theta\left(Y\sin\theta\partial_\theta\Phi\right)+\frac{Y}{\sin^2\theta}\partial_\psi^2\Phi+a^2Y\sin^2\theta\partial_t^2\Phi\,,\label{kgeqint1}
\end{align}
where we have defined 
\begin{equation}
Y=Y\left(r,\theta\right)\equiv\sqrt{\frac{G(r,\theta)}{F(r,\theta)}}\,.
\end{equation}
Then, we consider the following decomposition:
\begin{equation}
\Phi(t,r,\theta,\psi)\equiv e^{-i\omega t+im\psi}\,\Psi(r,\theta)\,,
\end{equation}
such that Eq.~\eqref{kgeqint1} can be rewritten as
\begin{equation}
\partial_r\left(\Delta Y\partial_r\Psi\right)+\partial_y\left[Y(1-y^2)\partial_y\Psi\right]+V(r,y,m,\omega)\Psi=0\,,\label{KGint2}
\end{equation}
where we have defined $y\equiv\cos\theta$. The effective potential $V(r,y,m,\omega)$ reads
\begin{align}
&V(r,y,m,\omega)\nonumber\\=&\,Y\Bigg[a^2\left(1-y^2\right)\mu^2Y-X\mu^2Y+\frac{X^2\omega^2}{\Delta}-\frac{m^2}{1-y^2}\nonumber\\&-a^2\left(1-y^2\right)\omega^2+\frac{a^2m^2}{\Delta}+\frac{2a}{\Delta}\left(\Delta-X\right)\omega m\Bigg]\,.\label{effeV}
\end{align}

Now, we investigate the necessary condition for Eq.~\eqref{KGint2} to be separable in terms of $r$ and $y$. First, according to the first two terms of Eq.~\eqref{KGint2}, a necessary condition is that $Y(r,y)$ must be a product of a function of $r$ and a function of $y$: 
\begin{equation}
Y(r,y)=Y_r(r)Y_y(y)\,.
\end{equation}
Furthermore, the asymptotic flatness condition requires that $Y(\infty,y)\rightarrow1$. This implies that $Y_y(y)=1$. Also, for the separability of Eq.~\eqref{KGint2}, the effective potential divided by $Y_r(r)$, that is, $V(r,y,m,\omega)/Y_r(r)$, is required to be written as a sum of a function of $r$ and a function of $y$ for any given $\omega$, $m$, and $\mu$. It can be seen that the $y^2\mu^2Y$ on the right hand side of Eq.~\eqref{effeV} violates this requirement, unless $Y=Y_r=\textrm{constant}$. This constant is then fixed to unity due to the asymptotic flatness condition. As a result, for the general stationary and axisymmetric black holes generated through NJA, the separability of the massive Klein-Gordon equation requires $F(r,\theta)=G(r,\theta)$. If this requirement is fulfilled, considering the ansatz $\Psi(r,y)=R(r)\Theta(y)$, the Klein-Gordon equation can be separated into an angular part and a radial part, which read
\begin{align}
\Bigg\{\partial_y\left[\left(1-y^2\right)\partial_y\right]&+\left(\omega^2-\mu^2\right)a^2y^2\nonumber\\&-\frac{m^2y^2}{1-y^2}-\left(m-a\omega\right)^2+C\Bigg\}\Theta(y)=0\,,
\end{align}
and
\begin{align}
\Bigg[\partial_r\left(\Delta\partial_r\right)+\frac{\left(\omega X-am\right)^2}{\Delta}-\left(X-a^2\right)\mu^2-C\Bigg]R(r)=0\,,
\end{align}
respectively, where $C$ is a separation constant.

Before closing this section, we would like to mention that the condition $Y=1$ for the separability of the Klein-Gordon equation can be relaxed when one considers a massless scalar field $(\mu=0)$. The massless Klein-Gordon equation can be separable as long as $Y=Y_r(r)$ and the function $Y_r(r)$ satisfies the asymptotic flatness condition ($Y_r(\infty)\rightarrow1$) for a physically viable solution.

\section{Comparison with other parametrized Kerr-like metrics}\label{sec.4}
The general stationary and axisymmetric spacetime obtained through NJA, i.e., Eq.~\eqref{NJAmetric} is described by three functions: $\Delta(r)$, $X(r)$, and $H(r,\theta)$, and in principle it can be used to parametrize possible deviations from the Kerr spacetime. Therefore, in this section we will compare the NJA metric \eqref{NJAmetric} with other parametrization approaches in the literature. 

\subsection{The Johannsen parametrized spacetime}
The first parametrized Kerr-like spacetime we are going to consider is the Johannsen parametrized spacetime \cite{Johannsen:2015pca}, whose spacetime metric is given by
\begin{align}
g_{\theta\theta}&=\tilde\Sigma\,,\qquad g_{rr}=\frac{\tilde\Sigma}{\Delta_0A_5(r)}\,,\nonumber\\
g_{tt}&=-\frac{\tilde\Sigma\left[\Delta_0-a^2A_2(r)^2\sin^2\theta\right]}{\Gamma^2}\,,\nonumber\\
g_{t\psi}&=-\frac{a\left[\left(r^2+a^2\right)A_1(r)A_2(r)-\Delta_0\right]\tilde\Sigma\sin^2\theta}{\Gamma^2}\,,\nonumber\\
g_{\psi\psi}&=\frac{\tilde\Sigma\sin^2\theta\left[\left(r^2+a^2\right)^2A_1(r)^2-a^2\Delta_0\sin^2\theta\right]}{\Gamma^2}\,,\label{Jps}
\end{align} 
where
\begin{align}
\Gamma&\equiv\left(r^2+a^2\right)A_1(r)-a^2A_2(r)\sin^2\theta\,,\nonumber\\
\tilde\Sigma&\equiv r^2+a^2\cos^2\theta+\tilde{f}(r)\,,\nonumber\\
\Delta_0&\equiv r^2-2Mr+a^2\,.\label{johna2}
\end{align}
In can be seen that the Johannsen parametrized spacetime is determined by four functions $A_1(r)$, $A_2(r)$, $A_5(r)$, and $\tilde{f}(r)$. This metric is featured by the existence of the Carter constant of motion, which implies the separability of the Hamilton-Jacobi equation and null geodesic equations \cite{Johannsen:2015pca}.

As we have mentioned previously, the NJA metric \eqref{NJAmetric} also allows for separable null geodesic equations \cite{Shaikh:2019fpu}. In addition, the Hamilton-Jacobi equation of the NJA metric is separable when $H(r,\theta)$ is an additively separable function. In fact, the two metrics \eqref{NJAmetric} and \eqref{Jps} are compatible if and only if
\begin{equation}
H(r,\theta)=\tilde\Sigma\,,\quad\textrm{and}\quad A_5(r)A_2(r)^2=1\,.\label{NJAjps}
\end{equation}
Furthermore, if Eq.~\eqref{NJAjps} is satisfied, it can be shown that
\begin{equation}
\Delta(r)=\frac{\Delta_0}{A_2^2}\,,\qquad X(r)=\left(r^2+a^2\right)\frac{A_1}{A_2}\,.
\end{equation}

It should be mentioned that the author of Ref.~\cite{Shaikh:2019fpu} has shown that the shadow of the NJA metric \eqref{NJAmetric} is completely determined by $\Delta(r)$ and $X(r)$ only. This is consistent with the results in Ref.~\cite{Johannsen:2015qca}, showing that the shadow of the Johannsen parametrized spacetime only depends on the deviation parameters in $A_1(r)$ and $A_2(r)$. 

\subsection{The KSZ parametrized spacetime}
In Ref.~\cite{Konoplya:2016jvv}, Konoplya, Rezzolla and Zhidenko proposed a general axisymmetric spacetime metric whose metric functions are expressed as generic functions of $r$ and $\theta$. We shall refer to this metric as KRZ spacetime. In Ref.~\cite{Konoplya:2018arm}, furthermore, the condition of the separability of the Klein-Gordon and the Hamilton-Jacobi equations for the KRZ parameterized spacetime has been investigated, and a subclass of KRZ spacetime is obtained (we then refer to it as KSZ spacetime). The KSZ parameterized spacetime is given by
\begin{align}
g_{tt}&=-\frac{N^2-W^2\sin^2\theta}{K^2}\,,\quad g_{t\psi}=-Wr\sin^2\theta\,,\nonumber\\
g_{\psi\psi}&=K^2r^2\sin^2\theta\,,\quad g_{\theta\theta}=r^2R_\Sigma(r)+a^2\cos^2\theta\,,\nonumber\\
g_{rr}&=\frac{g_{\theta\theta}R_B(r)^2}{r^2N^2}\,,\label{KSZmetric}
\end{align}
where 
\begin{align}
W&=\frac{a R_M(r)}{r^2R_\Sigma(r)+a^2\cos^2\theta}\,,\nonumber\\
N^2&=R_\Sigma(r)-\frac{R_M(r)}{r}+\frac{a^2}{r^2}\,,\nonumber\\
K^2&=\frac{r^2R_\Sigma^2+a^2R_\Sigma+a^2N^2\cos^2\theta}{r^2R_\Sigma(r)+a^2\cos^2\theta}+\frac{aW}{r}\,.
\end{align}
In Ref.~\cite{Konoplya:2018arm}, it has been proven that the KSZ spacetime allows for the separability of the Klein-Gordon equation and the Hamilton-Jacobi equation. Essentially, the KSZ spacetime is determined by three functions $R_\Sigma(r)$, $R_M(r)$, and $R_B(r)$. Actually, only two of them are independent since one can redefine the radial variable to fix one of these functions.  

The KSZ metric and the NJA metric \eqref{NJAmetric} are compatible if and only if
\begin{align}
R_\Sigma(r)&=\frac{X(r)-a^2}{r^2}\,,\quad R_M(r)=\frac{X(r)-\Delta(r)}{r}\,,\nonumber\\
R_B(r)&=1\,,\qquad H(r,\theta)=X(r)-a^2\sin^2\theta\,.\label{KSZnja}
\end{align}
Therefore, the functions $\Delta$ and $X$ completely determine $R_\Sigma$ and $R_M$. Also, according to Eq.~\eqref{DeltaX}, the last equation in Eq.~\eqref{KSZnja} implies $F(r,\theta)=G(r,\theta)$, which is just the necessary condition for the separability of the Klein-Gordon equation of the metric \eqref{NJAmetric} as we have proven in section~\ref{sec.3}.  Finally, since the third equation in Eq.~\eqref{KSZnja} turns out to be just a gauge choice, one can therefore conclude that the KSZ parameterized spacetime is a subclass of the metric \eqref{NJAmetric}. It should be also emphasized that if a metric belongs to the NJA metric and its Klein-Gordon equation is separable, then it must belong to the KSZ metric as well.

\subsection{The PK parameterized spacetime}
In Ref.~\cite{Papadopoulos:2018nvd}, Papadopoulos and Kokkotas proposed an innovative approach to construct the most general axisymmetric spacetimes which respect the preservation of the Carter constant and the asymptotic flatness condition. In the contravariant form, the metric tensor can be expressed as \cite{Papadopoulos:2018nvd}
\begin{align}
g^{tt}&=\frac{\mathcal{A}_5(r)+\mathcal{B}_5(y)}{\mathcal{A}_1(r)+\mathcal{B}_1(y)}\,,\quad g^{t\psi}=\frac{\mathcal{A}_4(r)+\mathcal{B}_4(y)}{\mathcal{A}_1(r)+\mathcal{B}_1(y)}\,,\nonumber\\
g^{\psi\psi}&=\frac{\mathcal{A}_3(r)+\mathcal{B}_3(y)}{\mathcal{A}_1(r)+\mathcal{B}_1(y)}\,,\quad g^{yy}=\frac{\mathcal{B}_2(y)}{\mathcal{A}_1(r)+\mathcal{B}_1(y)}\,,\nonumber\\
g^{rr}&=\frac{\mathcal{A}_2(r)}{\mathcal{A}_1(r)+\mathcal{B}_1(y)}\,,\label{PK}
\end{align}
where $\mathcal{A}_i(r)$ and $\mathcal{B}_i(y)$ are arbitrary functions. The metric \eqref{PK} is the most general axisymmetric metric for the preservation of the Carter constant and the asymptotic flatness condition in the spacetime. In Ref.~\cite{Papadopoulos:2018nvd}, it has been shown that the Johannsen metric given in Eqs.~\eqref{Jps} is just a subclass of the metric \eqref{PK}.

As we have mentioned, the separability of the Hamilton-Jacobi equation of the NJA metric \eqref{NJAmetric} requires that the function $H(r,y)$ is additively separable, i.e., $H(r,y)=H_1(r)+H_2(y)$. In fact, the NJA metric \eqref{NJAmetric} and the PK metric \eqref{PK} are compatible if and only if
\begin{align}
\mathcal{A}_1(r)&=H_1(r)\,,\qquad \mathcal{B}_1(y)=H_2(y)\,,\nonumber\\
\mathcal{A}_2(r)&=\Delta(r)\,,\qquad \mathcal{B}_2(y)=1-y^2\,,\nonumber\\
\mathcal{A}_3(r)&=-\frac{a^2}{\Delta(r)}\,,\qquad \mathcal{B}_3(y)=\frac{1}{1-y^2}\,,\nonumber\\
\mathcal{A}_4(r)&=-\frac{aX(r)}{\Delta(r)}\,,\qquad \mathcal{B}_4(y)=a\,,\nonumber\\
\mathcal{A}_5(r)&=-\frac{X(r)^2}{\Delta(r)}\,,\qquad \mathcal{B}_5(y)=a^2\left(1-y^2\right)\,.\label{NJAPK}
\end{align}
Therefore, the NJA metric \eqref{NJAmetric} is a subclass of the PK metric, subject to the condition $H(r,y)=H_1(r)+H_2(y)$. If the metric function $H(r,y)$ is not additively separable, the NJA metric does not belong to the PK metric and its Hamilton-Jacobi equation is not separable in general.

\begin{figure}[t]
\includegraphics[scale=0.24]{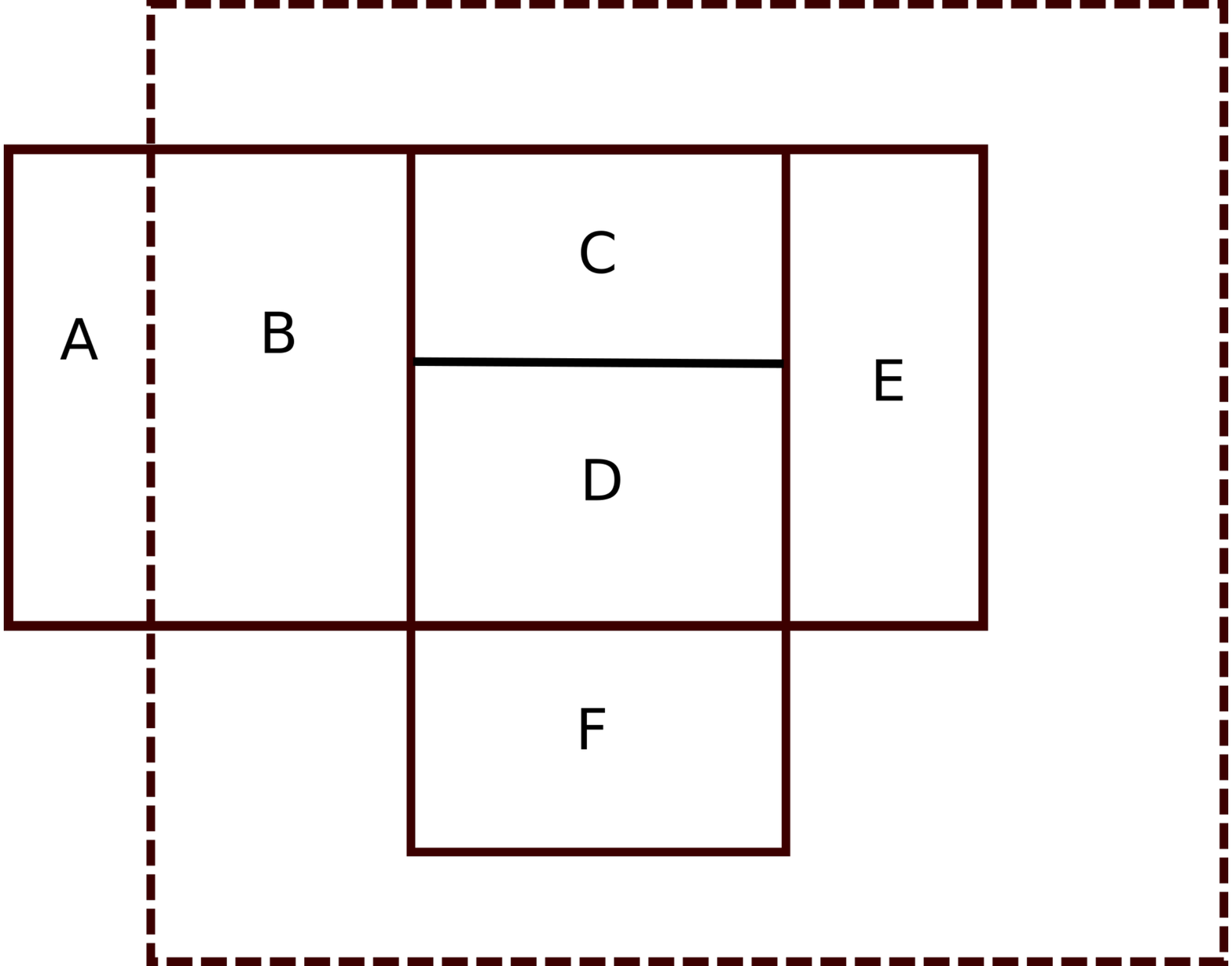}
\caption{\label{f3}This Venn diagram shows the relations among the sets of different metrics discussed in this paper. The NJA metric is given by the union A $\cup$ B $\cup$ C $\cup$ D. The Johannsen metric is depicted by the union C $\cup$ D $\cup$ E. The KSZ metric corresponds to the region D. The Carter's metric is depicted by the union D $\cup$ F. Finally, the PK metric is given by the region enclosed by the dashed rectangle.}
\end{figure}

It would be worth mentioning that there exists a set of PK metrics (the region B in Figure~\ref{f3}) which belongs to the NJA metric, but does not belong to the Johannsen metric. This particular set of metric contains an additively separable $H(r,y)$, but the function $H_2(y)$ is an arbitrary function of $y$ and cannot be expressed as $a^2y^2+\textrm{constant}$.

\subsection{Carter's metric}
Finally, we include the comparison of our metrics with the metric proposed by Carter in Ref.~\cite{Carter:1968ks} (we refer to it as Carter's metric). The Carter's metric allows the separability of the Hamilton-Jacobi equation and the analogues Schr\"odinger equation. The separability of the latter imposes a stronger restriction on the metric and the metric can be expressed as a simple algebraic form \cite{Carter:1968ks}. The metric can be written as
\begin{align}
g_{tt}&=\frac{\Delta_y(y)Q_r(r)^2-\Delta_r(r)Q_y(y)^2}{\mathcal{Z}(r,y)}\,,\nonumber\\g_{t\psi}&=\frac{\Delta_r(r)P_y(y)Q_y(y)-\Delta_y(y)P_r(r)Q_r(r)}{\mathcal{Z}(r,y)}\,,\nonumber\\
g_{\psi\psi}&=\frac{\Delta_y(y)P_r(r)^2-\Delta_r(r)P_y(y)^2}{\mathcal{Z}(r,y)}\,,\nonumber\\ g_{yy}&=\frac{\mathcal{Z}(r,y)}{\Delta_y(y)}\,,\quad
g_{rr}=\frac{\mathcal{Z}(r,y)}{\Delta_r(r)}\,,\label{CarterPK}
\end{align}
where $\Delta_r(r)$, $P_r(r)$, $Q_r(r)$, $\Delta_y(y)$, $P_y(y)$, and $Q_y(y)$ are arbitrary functions, subject to the condition that the function $\mathcal{Z}\equiv P_rQ_y-P_yQ_r$ is an additively separable function, i.e., $\mathcal{Z}(r,y)=\mathcal{Z}_1(r)+\mathcal{Z}_2(y)$.

First of all, it can be shown that the Carter's metric is a subclass of the PK metric according to the following allocations
\begin{align}
\mathcal{A}_1(r)&=\mathcal{Z}_1(r)\,,\qquad \mathcal{B}_1(y)=\mathcal{Z}_2(y)\,,\nonumber\\
\mathcal{A}_2(r)&=\Delta_r(r)\,,\qquad \mathcal{B}_2(y)=\Delta_y(y)\,,\nonumber\\
\mathcal{A}_3(r)&=-\frac{Q_r^2}{\Delta_r}\,,\qquad \mathcal{B}_3(y)=\frac{Q_y^2}{\Delta_y}\,,\nonumber\\
\mathcal{A}_4(r)&=-\frac{P_rQ_r}{\Delta_r}\,,\qquad \mathcal{B}_4(y)=\frac{P_yQ_y}{\Delta_y}\,,\nonumber\\
\mathcal{A}_5(r)&=-\frac{P_r^2}{\Delta_r}\,,\qquad \mathcal{B}_5(y)=\frac{P_y^2}{\Delta_y}\,.
\end{align}
According to Eqs.~\eqref{NJAPK} and the results in Ref.~\cite{Papadopoulos:2018nvd} (Eqs.~(14) in that paper), we find that the NJA metric and the Johannsen metric share the same $\mathcal{B}_2$, $\mathcal{B}_3$, $\mathcal{B}_4$, and $\mathcal{B}_5$ when they are expressed under the PK parameterization. This particular set of $\mathcal{B}_2$, $\mathcal{B}_3$, $\mathcal{B}_4$, and $\mathcal{B}_5$ is equivalent to the following choice of Carter's metric functions
\begin{equation}
\Delta_y=1-y^2\,,\quad P_y=a\left(1-y^2\right)\,,\quad Q_y=1\,.\label{dpqstandard}
\end{equation}
To proceed, we fix these functions with Eq.~\eqref{dpqstandard} and compare the Carter's metric with the KSZ metric \eqref{KSZmetric}. We find that after fixing $R_B(r)=1$, these two metrics are related through the allocations
\begin{align}
\Delta_r(r)&=r^2R_\Sigma(r)-rR_M(r)+a^2\,,\nonumber\\
P_r(r)&=r^2R_\Sigma(r)+a^2\,,\qquad Q_r(r)=a\,,\nonumber\\
\mathcal{Z}(r,y)&=r^2R_\Sigma(r)+a^2y^2\,.
\end{align}
Therefore, the KSZ metric is a subclass of the Carter's metric.

Furthermore, the NJA metric is compatible with the Carter's metric if and only if 
\begin{align}
\Delta_y&=1-y^2\,,\quad P_y=a\left(1-y^2\right)\,,\quad Q_y=1\,,\nonumber\\
\Delta_r&=\Delta(r)\,,\quad P_r=X(r)\,,\quad Q_r=a\,.
\end{align}
In this case, we have 
\begin{equation}
\mathcal{Z}(r,y)=H(r,y)=X(r)-a^2\left(1-y^2\right)\,,
\end{equation}
which reduces to the last equation of \eqref{KSZnja}. Therefore, the intersection of the NJA metric and the Carter's metric is exactly the KSZ metric in which the Klein-Gordon equation is separable.

Finally, if we relax the condition of Eq.~\eqref{dpqstandard} or allow $Q_r(r)$ to be a varying function of $r$, the resultant Carter's metrics would neither belong to the NJA metric nor the Johannsen metric.

\subsection{Discussions}

In Figure~\ref{f3}, we show the Venn diagram to illustrate the relations between the sets of NJA metric (A $\cup$ B $\cup$ C $\cup$ D), the Johannsen metric (C $\cup$ D $\cup$ E), the KSZ metric (the region D), the Carter's metric (D $\cup$ F), and the PK metric (enclosed by the dashed rectangle). Except for the region A where $H(r,y)$ is not additively separable, all the metrics represented in this Venn diagram allow for a separable Hamilton-Jacobi equation and they belong to the PK metric. The region E represents the Johannsen metrics whose $A_5(r)A_2(r)^2\ne1$, so they do not belong to the set of NJA metrics. The region B represents the NJA metrics whose metric function $H(r,y)$ is additively separable but cannot be written as $H_1(r)+a^2y^2$. The intersection of the sets of the NJA metrics and the Johannsen metrics is the region C $\cup$ D. The metrics in this intersection satisfy Eq.~\eqref{NJAjps}. The region D stands for the KSZ metric and it is a subclass of the Johannsen metric and the NJA metric. The metrics in this region allow for a separable Klein-Gordon equation. Finally, the Carter's metric is depicted in the region D $\cup$ F. The region $F$ corresponds to the metrics in which the condition \eqref{dpqstandard} is relaxed or $Q_r(r)$ is not a constant. The metrics in the region $F$ would neither belong to the NJA metric nor the Johannsen metric.  

\section{Conclusions}\label{sec.5}
In this work, we consider the general stationary and axisymmetric spacetime generated through NJA and study what the criteria the metric functions should satisfy in oder to guarantee the separability of the Klein-Gordon equation. In general, the original spacetime metric is described by three functions: $\Delta(r)$, $X(r)$, and $H(r,\theta)$. We have found that a relation between $X(r)$ and $H(r,\theta)$ should be imposed for the separability of the Klein-Gordon equation. 

Then, we have studied the relations between the NJA metric and four parameterized Kerr-like spacetimes in the literature. The first one is the Johannsen parameterized metric \cite{Johannsen:2015pca} which allows for the separability of the Hanilton-Jacobi equation. We have found the condition for the intersection of the Johannsen and the NJA metrics where the two metrics are compatible. The second metric is the KSZ metric \cite{Konoplya:2018arm} in which the Hamilton-Jacobi and the Klein-Gordon equations are both separable. We have shown that the set of the KSZ metric lies in the intersection of the sets of the NJA and the Johannsen metrics. The third metric is the PK metric \cite{Papadopoulos:2018nvd}, which is the most general axisymmetric metric for the preservation of the Carter constant and the asymptotic flatness condition in the spacetime. We have found that the metric function $H(r,y)$ in the NJA metric should be additively separable, otherwise it does not belong to the PK metric. Finally, the fourth metric that we have considered is the Carter's metric \cite{Carter:1968ks}. This metric allows for a separable Hamilton-Jacobi equation and the analogues Schr\"odinger equation. We have found that by adjusting properly the metric functions (the condition \eqref{dpqstandard} and $Q_r(r)=a$), the Carter's metric reduces to the KSZ metric. The Venn diagram in Figure~\ref{f3} illustrates the relations among these sets of metrics.

The separability of the Klein-Gordon equation in the general stationary and axisymmetric spacetime is known to be important in studying several physical problems, such as quasinormal modes and scattering problems around the rotating black hole. This investigation can be further extended to the separability condition for the Maxwell equations and gravitational perturbation equations. We leave these issues for our future works. 

\section*{Acknowledgments}
 
CYC and PC are supported by Ministry of Science and Technology (MOST), Taiwan, through No. 107-2119-M-002-005, Leung Center for Cosmology and Particle Astrophysics (LeCosPA) of National Taiwan University, and Taiwan National Center for Theoretical Sciences (NCTS). CYC is also supported by MOST, Taiwan through No. 108-2811-M-002-682. PC is in addition supported by US Department of Energy under Contract No. DE-AC03-76SF00515.

\end{document}